\documentclass[traditabstract,printer]{aa}
\usepackage{graphicx}
\usepackage{epsfig}
\usepackage{amssymb,amsmath}
\usepackage{natbib}
\bibpunct{(}{)}{;}{a}{}{,}
\usepackage{pslatex}
\usepackage{caption}
\usepackage{longtable}
\usepackage{hyperref}
\usepackage{gensymb}

\begin{document}

\title{Evolution of newborn rapidly rotating magnetars: effects of $R$-mode and fall-back accretion}

\author{Jie-Shuang Wang$^{1,2,3}$ and Zi-Gao Dai$^{1,2}$\thanks{E-mail: \href{dzg@nju.edu.cn}{dzg@nju.edu.cn}}}

\institute{
$^{1}$ School of Astronomy and Space Science, Nanjing University, Nanjing 210093, China\\
$^{2}$ Key Laboratory of Modern Astronomy and Astrophysics (Nanjing University),
Ministry of Education, Nanjing 210093, China\\
 $^{3}$Max-Planck-Institut f\"ur Kernphysik, Saupfercheckweg 1, D-69117 Heidelberg, Germany}


\authorrunning{Wang \& Dai}
\titlerunning{Effects of $r$-modes in accreting magnetars}

\begin{abstract}{
In this paper we investigate effects of the $r$-mode instability on a newborn rapidly-rotating
magnetar with fall-back accretion.
Such a magnetar could usually occur in core-collapse supernovae and gamma-ray bursts.
We find that the magnetar's spin and $r$-mode evolution are influenced by accretion. If the
magnetar is sufficiently spun up to a few milliseconds, gravitational radiation leads to the
growth of the $r$-mode amplitude significantly. The maximum $r$-mode amplitude reaches
an order $\sim 0.001$ when the
damping due to the growth of a toroidal magnetic field balances the growth of the $r$-mode
amplitude. If such a sufficiently spun-up magnetar was located at a distance
less than 1\,Mpc, then gravitational waves would be detectable by the Einstein Telescope
but would have an extremely low event rate.
However, if the spin-up is insufficient, the growth of the $r$-mode amplitude is
mainly due to the accretion torque. In this case, the maximum $r$-mode amplitude is of the order of
$\sim 10^{-6}-10^{-5}$.
}\end{abstract}


\maketitle

\section{Introduction}
The $r$-mode instability is of great astrophysical importance, since
it occurs in all rotating perfect fluid neutron stars \citep{Andersson1998,Friedman1998}.
This instability is driven by gravitational radiation via the Chandrasekhar-Friedman-Schutz
mechanism \citep{Chandrasekhar1970,Friedman1978}, which leads to rapid growth of
the $r$-mode amplitude, while viscosity acts to suppress the instability.
Given the dependence of gravitational wave emission on the neutron star spin
and the dependence of viscosity on the neutron star temperature, the critical angular
velocity--temperature curve dividing the $r$-mode stability from instability
(hereafter $r$-mode instability window) was obtained by \cite{Lindblom1998},
who also found that gravitational radiation of the $l=2$ current multipole is stronger
than the other multipoles. \cite{Owen1998} carried out a detailed study of the $r$-mode
instability in rapidly rotating neutron stars and calculated the gravitational
wave forms. Such gravitational wave signals might be detected at a distance of 20 Mpc by the
advanced Laser Interferometer Gravitational-wave Observatory (aLIGO).
Further studies were carried out to obtain the properties of the $r$-mode
instability especially for gravitational waves from rotating neutron stars
\citep{Andersson1999a,Kokkotas1999,Lindblom1999,Andersson2000,Bildsten2000,Lindblom2000a,
Lindblom2000b,Lindblom2001,Andersson2001,Arras2003,Bondarescu2007,Bondarescu2009,Yang2010,
Haskell2012,Alford2014a,Alford2014b,Alford2015,Chugunov2015,Haskell2015,Mytidis2015}
and rotating strange stars
\citep{Madsen1998,Madsen2000,Andersson2002,Zheng2006,Alford2014b,Dai2015}.

Neutron stars form during supernova (SN) explosions. Fall-back accretion could
happen after the core collapse of massive stars \citep{Michel1988,Chevalier1989}.
Owing to rotating progenitors, a disk is likely to form before fall-back materials are
accreted onto the neutron star surface \citep[e.g., we refer to][]{Piro2011}. Such a fall-back disk is
responsible for the jets of pulsars \citep{Blackman2004}, and
early afterglows of gamma-ray bursts \citep[GRBs;][]{Dai2012}. The $r$-mode instability of
neutron stars (with sub-Eddington accretion) in low-mass X-ray binary (LMXB)
systems was investigated \citep{Levin1999,Andersson1999b,Brown2000,Haskell2012,Mahmoodifar2013}.
The accreted materials in LMXBs spin up old neutron stars significantly and cause these
stars to enter the $r$-mode instability window again. Following this, the heating due to shear viscosity
affects the stellar thermal evolution while gravitational radiation
spins down the star. The effects of hypercritical advection-dominated accretion flows
without magnetic fields in close binary systems were investigated by \cite{Yoshida2000},
who found that hyperaccreting disks considerably increase the stellar angular momentum, which can
balance the spin-down due to gravitational radiation.

Newborn neutron stars are generally strongly magnetized. The average magnetic field
of isolated radio pulsars is $\sim 2\times 10^{12}$ G \citep{Arzoumanian2002}.
However, some astrophysical phenomena were driven by magnetars with
field strengths up to $10^{16}$ G, for example GRBs
\citep{Usov1992,Dai1998a,Dai1998b,Zhang2001,Thompson2004,Metzger2007,Dall'Osso2011}, soft gamma-ray repeaters
\citep[SGRs,][]{Thompson1995,Thompson1996,Vasisht1997,Kouveliotou1998} and some superluminous
supernovae \citep{Kasen2010,Woosley2010,Wang2015}. Based on the Australia Telescope
National Facility (ATNF) catalog\footnote{\url{http://www.atnf.csiro.au/people/pulsar/psrcat/}}
\citep{Manchester2005}, the highest magnetic field discovered so far is $\sim 2\times10^{15}$ G
in SGR 1806-20. Such an ultrastrong magnetic field plays an important role in $r$-mode evolution
\citep{Ho2000,Rezzolla2000,Rezzolla2001a,Rezzolla2001b,Lee2005,Cuofano2010,Cuofano2012,
Abbassi2012,Asai2015,Chugunov2015}.

Fall-back accretion of rapidly rotating magnetars could drive a subset of transients.
For example, the propeller effect during fall-back accretion of newborn millisecond magnetars
could directly power some broad-lined type Ib/c or IIP SNe \citep{Piro2011};
the early bumps of some GRB afterglows
also hint that their central engines are hyper-accreting millisecond magnetars \citep{Dai2012}.
On one hand, the $r$-mode could be unstable in these magnetars and affect the spin
evolution via gravitational radiation. On the other hand, both magnetic dipole radiation and
fall-back accretion can affect evolution of the stellar spin and $r$-mode amplitude.
In this paper, we explore the effects of the $r$-mode instability on evolution of
newborn hyper-accreting magnetars, and test whether or not gravitational waves
due to the $r$-mode are detectable. If detected, such gravitational
waves could provide an important probe for central engines of some SNe and GRBs.
It was found that magnetars could be gravitational wave sources due to toroidal
magnetic field-induced distortion of the neutron star \citep{Cutler2002,Dall'Osso2009} and the
bar-mode instability \citep{Corsi2009}. \cite{Piro2012} studied gravitational radiation
due to bar-mode-like oscillations in accreting magnetars. They found that for a magnetar
with radius $\sim20$ km gravitational radiation lasts between hundreds of and a few thousand seconds at
typical frequencies of $1000-2000$\,Hz \citep{Piro2012}. However, the $r$-mode instability
could dominate gravitational radiation if the saturation amplitude is large enough.
In this case, the frequencies of gravitational waves due to $r$-mode
instability are $\sim 100$-$1000$ Hz as is shown below, which are the most sensitive
frequencies of aLIGO \citep{Harry2010} and the Einstein Telescope (ET)
\citep{Hild2008}. Thus future detections of gravitational waves
could distinguish between these two kinds of oscillations.

In this paper, we investigate the effects of the $r$-mode instability on a newborn rapidly-rotating
magnetar with fall-back accretion. In section 2, we present some basic equations
of the $r$-mode instability and hyper-accreting disks in magnetars and then we obtain
evolution equations of the stellar spin and $r$-mode amplitude. The numerical and semi-analytical
results are presented in section 3. In section 4, we estimate the detectability of gravitational
waves due to the $r$-mode instability. Our conclusions and brief discussions are given in section 5.

\section{Basic equations of $R$-mode instability}

\subsection{Gravitational radiation and viscosities}
The $r$-mode instability is generically driven by gravitational radiation and
dissipated by viscosities. The typical timescales of gravitational radiation ($t_{\rm gw}$),
shear viscosity ($t_{\rm sv}$), and bulk viscosity ($t_{\rm bv}$) depend on the stellar
equation of state (EOS). A polytropic EOS of $N=0$ or 1 is usually used, while the resulting
timescales for these two polytropic EOSs differ only by roughly a factor of two (we refer to a review
by \cite{Andersson2001} and references therein). The effects of more realistic EOSs
have been tested by \cite{Papazoglou2015}, whose results show only small differences
from those of the $N=1$ polytropic EOS. Following \cite{Alford2012a,Alford2012},
we consider the Akmal-Pandharipande-Ravenhall (APR) EOS \citep{Akmal1998}, since
this EOS gives a better approximation of the equilibrium configuration than the polytropic
EOSs for $r$-mode oscillations.

The gravitational radiation timescale is given by \cite{Lindblom1998} and \cite{Owen1998},
\begin{eqnarray}
\frac{1}{t_{\rm gw}} & = & -\frac{32\pi G
\Omega^{2l+2}}{c^{2l+3}}\frac{(l-1)^{2l}}{[(2l+1)!!]^2}\nonumber \\ & & \times \left(\frac{l+2}{l+1}\right)^{2l+2}
\int_0^{R}\rho(r) r^{2l+2} dr,
\end{eqnarray}
where $G$ is the gravitational constant. The strongest gravitational radiation reaction
comes from the $l=m=2$ mode \citep{Lindblom1998,Owen1998},
so we obtain
\begin{eqnarray}
t_{\rm{gw}}\approx-24 R_{11.5}^{-4}M_{1.4}^{-1}\nu_{3}^{-6}~{\rm s},\label{tgw}
\end{eqnarray}
where the radius of the magnetar is $R=11.5R_{11.5}$\,km, the mass is $M=1.4M_{\odot}\times M_{1.4}$,
and $\nu=\Omega/2 \pi=10^3\nu_{3}$\,Hz is the rotational frequency of the magnetar.

Usually very simple models of shear and bulk viscosities are adopted. However,
these models are inconsistent with observations of LMXBs even if taking into account the
effect of an Ekman layer \citep{Alford2014a}, and thus reevaluations of the
relevant physics are needed \citep{Ho2011,Haskell2012,Gusakov2014,Haskell2015}.
Here we consider more recent and complex models to calculate the dissipation
timescales of viscosities. The shear viscosity is found to be contributed by
electron scattering \citep{Shternin2008,Shternin2013}.
Following \cite{Alford2012a,Alford2012}, we adopt
this electron scattering model to calculate the dissipation timescale of shear
viscosity in this paper, which is
\begin{eqnarray}
t_{\rm sv}=1.4\times10^8M_{1.4}R_{11.5}^{-1}T_9^{5/3}\,{\rm s},
\end{eqnarray}
where $T=T_910^9$\,K is the core temperature of the neutron star.

The bulk viscosity is more important, since it is the dominant dissipation mechanism for
newly-born neutron stars. Most previous studies have focused
on the effects of the bulk viscosity from the linear perturbation theory. However, as the $r$-mode
amplitude ($\alpha$) grows large enough $\alpha\gtrsim O(1)$, the bulk viscosity is
significantly enhanced due to the suprathermal effect \citep{Alford2012}. This effect
provides an alternative mechanism to saturate the $r$-mode amplitude with the
order $O(1)$ to $O(10)$ \citep{Alford2012}. In this case, the dissipation timescale of the bulk
viscosity ($t_{\rm bv}$) also depends on the $r$-mode amplitude $\alpha$. For a neutron star
dominated by the modified-URCA processes, this time-scale is expressed as
\citep[for more details, we refer to][]{Alford2012},
\begin{eqnarray}
\frac{1}{t_{\rm bv}}  =  1.4\times10^{-10}R_{11.5}^5M_{1.4}^{-1}T_9^6 \nu_3^2(1+284.5R_{11.5}^4\nu_3^4T_9^{-2}\alpha^2+
\nonumber\\  3.16\times10^4R_{11.5}^8\nu_3^8T_9^{-4}\alpha^4+
1.08\times10^6R_{11.5}^{12}\nu_3^{12}T_9^{-6}\alpha^6)\,{\rm s}^{-1}.\label{tbv}
\end{eqnarray}
The first term on the right-hand side, namely $t_{\rm bv}^{-1}=1.4\times10^{-10}R_{11.5}^5M_{1.4}^{-1}T_9^6 \nu_3^2$,
is consistent with the results
from the linear perturbation theory in previous studies \citep{Lindblom1998,Owen1998}.

\subsection{Effect of magnetic field}
In rotating stars, the Coriolis force ($F_{\rm c}$) provides the restoring force.
However, in strongly-magnetized neutron stars, the magnetic force ($F_{B}$) also acts
on the fluid motion. The effects of a magnetic field have been studied widely
\citep{Ho2000,Rezzolla2000,Rezzolla2001a,Rezzolla2001b,Lee2005,Cuofano2010,Cuofano2012,
Abbassi2012,Asai2015,Chugunov2015}. Differential rotation might be induced, and thus transfer
energy to produce a toroidal magnetic field
\citep{Rezzolla2000,Rezzolla2001a,Rezzolla2001b}. We consider the damping timescale due to
the generation of a toroidal magnetic field following \cite{Cuofano2010,Cuofano2012},
that is,
\begin{eqnarray}
t_{\rm B,t}\approx 5.8 R_{11.5}^{-1}M_{1.4}B_{15}^{-1}B_{\rm t,14}^{-1}\nu_{3}
~{\rm s},\label{tbt}
\end{eqnarray}
where $B=B_{15}10^{15}$\,G is the surface magnetic field of the magnetar and
$B_{\rm t}=B_{\rm t,14}10^{14}$\,G is the volume-averaged toroidal magnetic field, which
follows from
\begin{eqnarray}
\dfrac{d B_{\rm t}}{d t}\approx \left(\dfrac{4}{3\pi}\right)^{1/2}B\alpha^2\Omega.\label{tbevo}
\end{eqnarray}
A toroidal magnetic field in the interior can deform the neutron star to an ellipsoid, which leads
to gravitational radiation and meanwhile spins down the star by a torque of
$N_{\rm B,gw}=-I\Omega/t_{\rm B,gw}$, where $t_{\rm B,gw}$ is the timescale of gravitational radiation
related to this torque, $I=\tilde{I} M R^2$ is the moment of
inertia, and $\tilde{I}={8\pi\over3MR^2}\int_0^R\rho r^4dr$. It is a non-trivial problem that
the gravitational radiation from such a magnetically-distorted neutron star depends on details of the
magnetic field distribution in the stellar interior, as discussed by many authors
\citep{Cutler2002,Dall'Osso2009,Corsi2009,Cuofano2012,Dall'Osso2015}. If the
inclination angle between the magnetic dipole moment and the rotation axis is
$\chi=\pi/2$, the time-varying mass quadrupole of the neutron star and consequently
the gravitational radiation efficiency are maximized; otherwise, the gravitational
radiation efficiency should be smaller. If a
neutron star has an initial temperature $T_0=10^{10}$\,K and cools down via the modified URCA
process, as found by \cite{Dall'Osso2009}, then the inclination angle evolves following
\begin{eqnarray}
\frac{\mbox{sin}^2 \chi}{1+3\mbox{cos}^2 \chi} = \frac{\mbox{sin}^2 \chi_i}{1+3\mbox{cos}^2 \chi_i} \left(\frac{t}{30\,{\rm s}} +1
\right)^{240/C},
\end{eqnarray}
where $\chi_i$ is the initial inclination angle, which is usually small, and
 $C\sim2.3\times10^{-4}B_{\rm t,14}^2M_{1.4}^{-1}\nu_{3}^{-2}$.
Thus the magnetic dipole moment will be orthogonal to the rotation
axis in less than $\sim1$\,ms when $\chi_i=1\degree$. For the APR equation of state,
we obtain $\tilde{I}=0.283$ for a neutron star with mass $1.4M_{\odot}$ and radius
$R=11.5$\,km \citep{Owen1998}. The timescale $t_{\rm B,gw}$ with $\chi=\pi/2$
reads \citep{Cuofano2012}
\begin{eqnarray}
t_{\rm B,gw}\approx 3.8\times10^{14} M_{1.4}^{-1}R_{11.5}^{-2}B_{\rm t,14}^{-4}\nu_{3}^{-4}
~{\rm s}.\label{tbgw}
\end{eqnarray}

The surface magnetic field also slows down the magnetar. For simplicity, the magnetic dipole radiation
model is adopted. The torque due to magnetic dipole radiation and corresponding timescale are
then given by
\begin{eqnarray}
N_{\rm dip}\equiv -\frac{\mu^2\Omega^3\sin^2\chi}{6c^3}=-{\mu^2\over6R_{\rm L}^3},
\end{eqnarray}
and
\begin{eqnarray}
t_{\rm dip}\equiv I\Omega/N_{\rm dip}=-1.4\times10^3 B_{15}^{-2} \nu_{3}^{-2}\,{\rm s},
\end{eqnarray}
where $t_{\rm dip}$ is the typical timescale due to magnetic dipole
radiation and we set $\chi={\pi \over 2}$. $R_{\rm L}$ is the radius of the light cylinder
\begin{eqnarray}
R_{\rm L}=\frac{c}{\Omega}\approx48\nu_{3}^{-1}\,{\rm km},\label{rl}
\end{eqnarray}
where $c$ is the speed of light. The matter outside $R_{\rm L}$ no longer
interacts with the magnetar.

\subsection{Hypercritical accretion of a magnetar}

Previous studies of the $r$-mode instability in accretion neutron stars have focussed on LMXB
systems \citep{Andersson1999b,Brown2000,Yoshida2000,Bondarescu2007}.
Here we consider hypercritical fall-back accretion of newborn rapidly rotating  magnetars,
where the accretion torque can be either positive or negative. Following \cite{MacFadyen2001},
\cite{Zhang2008} and \cite{Piro2012}, we parameterize the fall-back accretion rate as
\begin{eqnarray}
\dot{M}=(\dot{M}_{\rm early}^{-1}+\dot{M}_{\rm late}^{-1})^{-1},
\end{eqnarray}
where
\begin{eqnarray}
\dot{M}_{\rm early}=10^{-3}\eta t^{1/2}M_\odot\,{\rm s}^{-1},\label{accearly}
\end{eqnarray}
and
\begin{eqnarray}
\dot{M}_{\rm late}=10^{-3}\eta t_0^{13/6}t^{-5/3}M_\odot\,{\rm
s}^{-1},\label{acclate}
\end{eqnarray}
where $\eta\approx0.01-10$ is a factor which accounts for explosion energies, $t$ is
in units of seconds and $t_0\sim10^2-10^3$ s is the typical time at which the
accretion rate starts to decrease \citep{MacFadyen2001}. At late times, the accretion
rate decreases as $\dot M\propto t^{-5/3}$ \citep{Chevalier1989}.
We thus obtain the total baryonic mass of the neutron star at an arbitrary time $t$,
\begin{eqnarray}
M_b(t)=M_{b,i}+\int_0^t\dot M_b dt,
\end{eqnarray}
where $M_{b,i}$ is the initial baryonic mass of the magnetar.
A fraction of the accreted mass will turn into the binding energy that will be radiated away
in the form of neutrinos. Hence, the gravitational mass ($M$) is more applicable
\citep{Lattimer2001},
\begin{eqnarray}
M=M_b\left(t\right)\left[1+\frac{3}{5}\frac{GM_b\left(t\right)}{R c^2}\right]^{-1}.
\end{eqnarray}
We take $R$ to be a constant in this paper. In reality, the radius $R$ changes especially when
the mass grows very large \citep{Lattimer2001}. However, a small change of $R$
cannot affect our results significantly, since the timescales acting on the $r$-mode mainly
depend on the core temperature and the spin.

In order to parameterize the torques due to accretion and magnetic dipole radiation,
we introduce two useful radii. The first one is the co-rotation radius where the Keplerian
angular velocity of the disk equals the angular velocity of the neutron star $\Omega$,
\begin{eqnarray}
r_{\rm c}=\left(\frac{GM}{\Omega^2}\right)^{1/3}\approx16.5M_{1.4}^{1/3}\nu_3^{-2/3}\,{\rm km}.\label{rc}
\end{eqnarray}
The second one is the magnetospheric (Alfv\'{e}n) radius where the ram pressure of the
fall-back materials balances the magnetic pressure,
\begin{eqnarray}
r_{\rm m}=\left(\frac{\mu^4}{GM\dot M_b^2}\right)^{1/7}
\approx31.2B_{15}^{4/7}R_{11.5}^{12/7}M_{1.4}^{-1/7}\dot M_{-3}^{-2/7}\,{\rm km},\label{rm}
\end{eqnarray}
where $\mu=B R^3$ is the magnetic dipole moment  and $\dot M_{-3}=\dot M_b/(10^{-3}M_{\odot}/$s).

Usually the fastness parameter $\omega$ is defined as the ratio of the stellar angular velocity
and the Keplerian angular velocity at the magnetospheric radius $r_{\rm m}$
\citep[we refer to e.g., ][]{Elsner1977,Ghosh1979},
\begin{eqnarray}
\omega & = & \frac{\Omega}{\sqrt{GM/r_{\rm m}^3}}=\left(\frac{r_{\rm m}}{r_{\rm c}}\right)^{3/2}\nonumber \\
& \approx & 2.6B_{15}^{6/7}R_{11.5}^{18/7}M_{1.4}^{-5/7}\dot M_{-3}^{-3/7}\nu_{3}.\label{omega}
\end{eqnarray}

Using these parameters, we can parameterize the accretion torque and corresponding timescale
through
\begin{eqnarray}
&N_{\rm acc} \equiv\dot M_b \sqrt{GM r_{\rm m}}n(\omega)=n(\omega)\mu^2/ r_{\rm m}^3,\label{tacc},
\end{eqnarray}
and
\begin{eqnarray}
&t_{\rm acc}\equiv I\Omega/N_{\rm acc}=180\nu_{3}\dot M_{-3}^{-1}({r_{\rm m}/18\,{\rm km}})^{1/2}n(\omega)^{-1}\,{\rm s}.
\end{eqnarray}
If $r_{\rm m}<r_{\rm c}$, the accreted matter transfers its angular momentum to
the magnetar, and thus provides a positive torque. Otherwise, the fall-back matter
acts as a propeller and offers a negative torque. A detailed picture of accretion
onto magnetic stars has been elucidated by many authors, for example \cite{Pringle1972},
\cite{Illarionov1975}, \cite{Ghosh1979},
\cite{Aly1980}, \cite{Eksi2005}, and \cite{Dai2012}. The expression of $n(\omega)$
differs slightly in different models. For simplification, we here adopt $n(\omega)=1-\omega$,
as in \cite{Piro2011}. This form of $n(\omega)$ includes the main properties of an
interaction between the accretion disk and the magnetar. Furthermore, it has two advantages;
$n(\omega)$ is continuous for all $\omega$ and becomes zero when $r_{\rm m}=r_{\rm c}$.

\subsection{Evolution of the $r$-mode}
We quantify the $r$-mode model based on \cite{Owen1998}.
The total angular momentum $J$ is given by
\begin{eqnarray}
J=I \Omega +J_{\rm c},\label{J}
\end{eqnarray}
where $J_{\rm c}$ is the canonical angular momentum of the $r$-mode
\citep{Owen1998},
\begin{eqnarray}
J_{\rm c}=-\frac{3}{2} \alpha ^2 \tilde{J} M R^2 \Omega,\label{Jc}
\end{eqnarray}
where $\tilde{J}=\int_0^R\rho r^6dr/(MR^4)$  and for a neutron star with the
APR EOS, whose mass is $1.4M_{\odot}$ and radius is $R=11.5$\,km, we have
$\tilde{J}=1.81\times 10^{-2}$ \citep{Owen1998}. The canonical energy of the $r$-mode is given by \citep{Owen1998},
\begin{eqnarray}
E_{\rm c}=\frac{1}{2} \alpha ^2 \tilde{J} M R^2 \Omega ^2.\label{Ec}
\end{eqnarray}

We obtain evolution of the $r$-mode using the time derivative of $J$ and $E_{\rm c}$
\citep{Owen1998,Andersson2001,Cuofano2012},
\begin{eqnarray}
\frac{d J}{dt}=\frac{3 \alpha ^2 \tilde{J} M R^2 \Omega }{t_{\rm gw}}+N_{\rm acc}+
N_{\rm dip}+N_{\rm B,gw},\label{dJ}\\
\frac{d E_{\rm c}}{dt}=-2 E_{\rm c} \left(\frac{1}{t_{\rm gw}}+
\frac{1}{t_{\rm sv}}+\frac{1}{t_{\rm bv}}+\frac{1}{t_{\rm B,t}}\right),\label{dE}
\end{eqnarray}
where the torques due to accretion ($N_{\rm acc}$) and magnetic dipole radiation
($N_{\rm dip}$) affect the total stellar angular momentum together.

Substituting equations (\ref{J}), (\ref{Jc}),
and (\ref{Ec}) into equations (\ref{dJ}) and (\ref{dE}), we obtain the evolution equations
of $\Omega$ and $\alpha$,
\begin{eqnarray}
{1\over\Omega}\frac{d\Omega }{dt} = -\frac{\dot M_b }{ A_{+}M}
+\frac{1}{A_{+}}\left(\frac{1}{t_{\rm dip}}+\frac{1}{t_{\rm acc}}+\frac{1}{t_{\rm B,gw}} \right)
\nonumber\\ -\frac{3 \alpha ^2 \tilde{J}}{\tilde{I}A_{+}}
\left(\frac{1}{t_{\rm sv}}+\frac{1}{t_{\rm bv}}+\frac{1}{t_{\rm B,t}}\right),\label{oevo}
\end{eqnarray}
and
\begin{eqnarray}
{1\over\alpha}\frac{d\alpha }{dt}  =  \frac{\dot M_b A_{-}}{2 M A_{+}}
-\frac{A_{-}}{ A_{+}}\left(\frac{1}{t_{\rm sv}}+\frac{1}{t_{\rm bv}}+\frac{1}{t_{\rm B,t}}\right)
\nonumber\\
-\frac{1}{ A_{+}}\left(\frac{1}{t_{\rm dip}}+\frac{1}{t_{\rm acc}}+\frac{1}{t_{\rm B,gw}} \right)-\frac{1 }{t_{\rm gw}}\label{aevo},
\end{eqnarray}
where $A_{\pm}=1\pm 3 \alpha ^2 \tilde{J}/2\tilde{I}$.
We see that accretion and magnetic dipole radiation affect the amplitude and spin evolution simultaneously.
The magnetic braking increases the $r$-mode amplitude while it spins down the magnetar. The effect of accretion
is slightly complicated, since accretion can either spin up or down
the magnetar while it can also either decrease or increase the $r$-mode amplitude.

In order to solve the above evolution equations, we must also know how the interior
temperature of the magnetar evolves. The fall-back accretion will start
$\sim 100$\,s later after the stellar birth \citep{MacFadyen2001,Dai2012}.
Meanwhile, the proto-neutron star
will become a neutron star $\sim 50$\,s later after its birth, and at this time the neutron star is
transparent to neutrinos \citep{Ferrari2003}. Thus we consider a hot young isothermal
neutron star with an initial temperature of $T_0=10^{10}$\,K. The newborn
neutron star is then cooled down by the modified-URCA processes, whose luminosity is
$L_{\nu}\approx3.5\times10^{39}T_9^8$ erg/s \citep{Shapiro1983,Alford2014b}.
The direct URCA processes might set in when the star grows heavy enough. But
since we only consider the time (the first $\sim10^4$\,s) when gravitational radiation is
significant, the direct URCA processes do not significantly affect our results.
We also take into account the heating effect due to shear viscosity
in this paper, namely, $H_{\rm v}=2E_{\rm c}/t_{\rm sv}$. We do not consider
the heating effect of accreted materials, since these
materials have been cooled down by neutrinos in the accretion disk \citep{Kohri2002,Zhang2010}
and the accretion column \citep{Piro2011}. The temperature evolution is then given by
\begin{eqnarray}
\frac{dT}{dt}=\frac{H_{\rm v}-L_{\nu}}{C_V},\label{Tevo}
\end{eqnarray}
where $C_V\approx7.0\times10^{38} T_9$ erg K$^{-1}$ \citep{Shapiro1983} is
the heat capacity of the neutron star.

\section{Results}

Integrating equations  (\ref{tbevo}), (\ref{oevo}), (\ref{aevo}) and (\ref{Tevo}),
we can obtain evolution of the $r$-mode amplitude, spin, and temperature.
The initial spin is of the order of one millisecond, but since accretion spins up the star near
the mass-shedding limit, the effect of the initial spin is less important than that
of accretion. Therefore, we mainly focus on effects of accretion and initialize the
spin at $P_0=1/\nu_0=3$\,ms. The initial baryonic mass and radius of the magnetar in
our model are taken to be $M_{b,i}=1.4\,M_\odot$ and $R=11.5$\,km, and thus the
gravitational mass is $M_i\simeq1.26\,M_\odot$. The initial amplitude of the $r$-mode is fixed at $10^{-8}$.

\begin{figure}
\begin{center}
\includegraphics[width=0.5\textwidth]{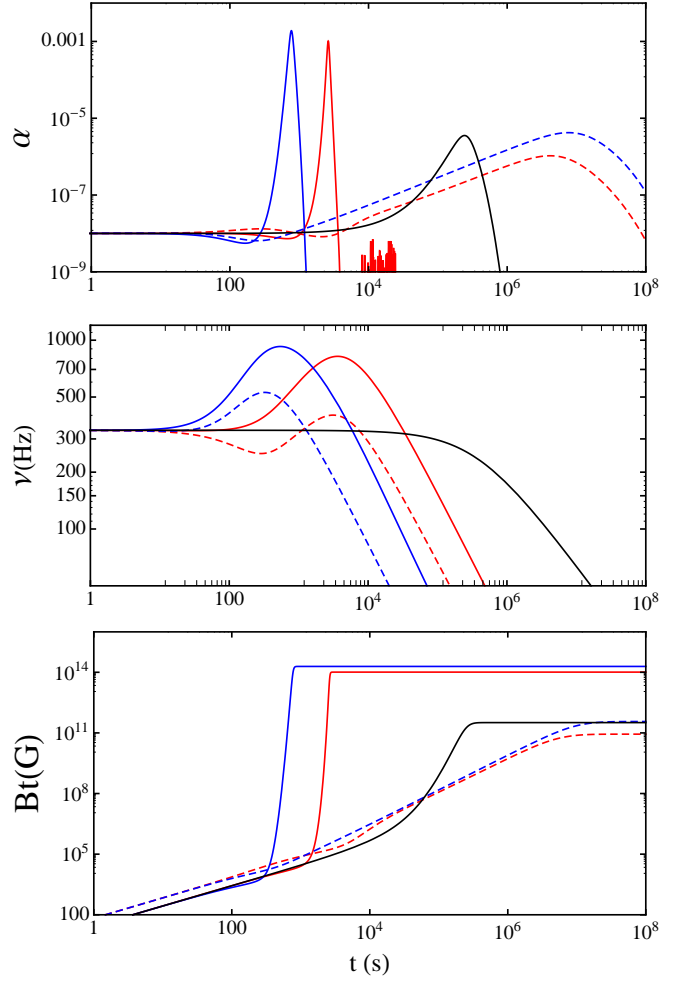}
\caption{The r-mode amplitude ($\alpha$), spin ($\nu$) and
toroidal magnetic field ($B_{\rm t}$) as functions of time. The black solid line is plotted
for $B=2\times10^{14}$\,G and $\eta=0$ . The red solid or dashed lines are obtained
in the case of $\eta=0.01$ and $t_0=3000$\,s for $B=2\times10^{14}$\,G or
$B=5\times10^{14}$\,G. The blue solid or dashed lines are obtained in the case
of $\eta=0.1$ and $t_0=300$\,s for $B=2\times10^{14}$\,G or
$B=5\times10^{14}$\,G.
\label{fig1}}
\end{center}
\end{figure}

Fig.\,\ref{fig1} gives four examples of evolution of the $r$-mode amplitude ($\alpha$),
spin ($\nu$) and toroidal magnetic field ($B_{\rm t}$). We study two specific accretion rates;
$\eta=0.01$ with $t_0=3000$\,s (red lines) and $\eta=0.1$ with $t_0=300$\,s (blue lines).
The magnetic field is typically chosen to be $B=2\times10^{14}$\,G (solid lines) and
$B=5\times10^{14}$\,G (dashed lines). For comparison, we also study the case without accretion
for $B=2\times10^{14}$\,G (black solid line). It is shown that the combination
of accretion and magnetic fields can significantly affect the $r$-mode amplitude evolution.
If the accretion can spin up the neutron star significantly, the amplitude grows quickly to the
maximum value of $\sim 0.001$. Otherwise, the growth speed might be considerably suppressed.
In the following, we use a semi-analytical method to analyze the amplitude and spin evolution.

To find which mechanism dominates the evolution, we compare the
terms in the right hand side of equation\,(\ref{oevo}) or (\ref{aevo}). We take the pre-factor
$A_{\pm}\approx1$, since the amplitude is very small in an early stage. Since
the viscosity terms and the gravitational wave terms due to magnetic deformation are very small,
we mainly consider the other terms. Firstly, we compare the accretion torque with the first term
in equation\,(\ref{oevo}), that is,
\begin{eqnarray}
\left|{M\over\dot M_b t_{\rm acc}}\right|
\approx 11B_{15}^{2/7}R_{11.5}^{-8/7}M_{1.4}^{3/7}\dot M_{-3}^{-1/7} \nu_{3}^{-1}|n(\omega)|.
\end{eqnarray}
Generally, $\nu<10^3$\,Hz holds in most of the evolution time (Fig.\,\ref{fig1})
and $n(\omega)$ is in the range of -0.5 to 0.5, so we treat with the accretion
torque as the dominant term over the first term.

The torque due to magnetic dipole radiation is smaller than the
accretion torque, since the magnetospheric radius is generally much smaller than
the light cylinder radius and $n(\omega)\in[-0.5,0.5]$ in the amplitude growing stage so that
\begin{eqnarray}
\left|{t_{\rm dip}\over t_{\rm acc}}\right|= \left|6\left({ R_{\rm L}\over r_{\rm m}}\right)^3 n(\omega)\right|>1.
\end{eqnarray}

The toroidal magnetic field damping term is comparable to the accretion term, if the
toroidal magnetic field grows strong enough, since we have
\begin{eqnarray}
{t_{\rm B,t}\over t_{\rm acc}}\approx 0.04 B_{\rm t,14}^{-1}
B_{15}^{-5/7}R_{11.5}^{-1/7}M_{1.4}^{13/14}\dot M_{-3}^{6/7} n(\omega).
\label{accovergw}
\end{eqnarray}
However, for the spin evolution (equation \ref{oevo}), there is a pre-factor
${3 \alpha ^2 \tilde{J}}/{\tilde{I}A_{+}}$. Considering
in this paper (Fig.\,\ref{fig1}) that the amplitude is generally much smaller than
0.2, we can assume that the accretion term dominates the spin evolution.

Adopting equations (\ref{accearly}) and (\ref{acclate}) and
considering only the accretion term in the right-hand side of equation (\ref{oevo}),
we obtain an approximate spin evolution,
\begin{eqnarray}
\nu\approx \left\{
\begin{array}{ll}
\nu_0+3.7n(\omega)\eta t^{3/2},~~~~{\rm if}~~t\lesssim t_0,\\
\nu_c+8.3n(\omega)t_0^{3/2}\eta[1-(t_0/t)^{2/3}],~~{\rm if}~~t\gtrsim t_0, \label{oad}
\end{array}
\right.
\end{eqnarray}
where $\nu_{c}=\nu_0+3.7n(\omega)\eta t_0^{3/2}$. Substituting equation\,(\ref{oad})
into equation\,(\ref{omega}), we find that $n(\omega)=0$ happens at
$t=1.5B_{15}^4 (\nu_{0,333})^{14/3}\eta^{-2}$\,s or
$t=110\eta^{3/5}B_{15}^{-6/5}(\nu_{c,800}^{-7/5})(t_{0,100})^{13/10}$\,s, where
$\nu_{0,333}=\nu_0/333$\,Hz, $\nu_{c,800}=\nu_c/800$\,Hz and $t_{0,100}=t_0/100$\,s.
The second solution ($t\sim 110\eta^{3/5}(t_{0,100})^{13/10}$\,s) corresponds to the
time when the maximum spin is reached, which hints that before $\sim t_0$ the spin is
increasing. This analytical expression for $\nu$ is consistent with the result shown in
Fig.\,\ref{fig1}.

The amplitude evolution begins to be affected by the gravitational wave emission when
\begin{eqnarray}
{|-t_{\rm gw}|\over t_{\rm acc}}\approx0.2
B_{15}^{2/7}R_{11.5}^{-36/7}M_{1.4}^{-11/7}\dot M_{-3}^{6/7} \nu_{3}^{-7}n(\omega)>1.
\label{accovergw}
\end{eqnarray}
This means that gravitational radiation dominates when
$\nu>794B^{2/49}_{15}\dot M^{6/49}_{-3}|n(\omega)|^{1/7}$\,Hz.
Approximately, it is required that $\nu_{c}>794B^{2/49}_{15}\dot M^{6/49}_{-3}|n(\omega)|^{1/7}$, which is roughly $3.7n(\omega)\eta t_0^{3/2}>794B^{2/49}_{15}
\eta^{6/49}t_0^{3/49}|n(\omega)|^{1/7}-\nu_0$.

\begin{figure}
\begin{center}
\includegraphics[width=0.5\textwidth]{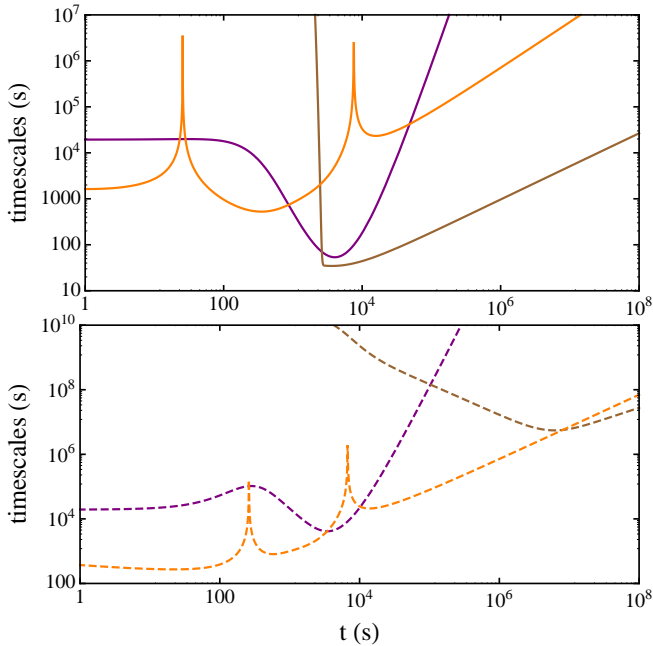}
\caption{The evolution of timescales $-t_{\rm gw}$ (purple lines), $t_{\rm B,t}$
(brown lines) and $t_{\rm acc}$ (orange lines) in the cases with $\eta=0.01$ and $t_0=3000$\,s.
The solid lines correspond to the case of $B=2\times10^{14}$\,G and dashed lines to
$B=5\times10^{14}$\,G.
\label{fig2}}
\end{center}
\end{figure}

We also numerically show that the timescales $-t_{\rm gw}$ (purple lines), $t_{\rm B,t}$
(brown lines) and $t_{\rm acc}$ (orange lines) evolve with time in Fig.\,\ref{fig2}. The
accretion rate is fixed to be $\eta=0.01$ at $t_0=3000$\,s. The solid lines correspond to the
case of $B=2\times10^{14}$\,G and dashed lines to $B=5\times10^{14}$\,G. The sharp
transitions happen when $\omega\rightarrow1$, which leads to $t_{\rm acc}\rightarrow\infty$.

The amplitude is suppressed by the positive accretion torque before $\sim t_0$.
After that, if the spin of the neutron star increases significantly enough, which
leads to gravitational radiation dominating the $r$-mode amplitude
($|{t_{\rm gw}|< |t_{\rm acc}}|$), rapid growth of the amplitude occurs
(Fig.\,\ref{fig1}). To calculate this rapid growth, we only consider the contribution
of gravitational radiation in equation\,(\ref{aevo}), and obtain
\begin{eqnarray}
\gamma\equiv\ln\left(\frac{\alpha}{\alpha_0}\right)\approx {1\over21}\int_0^t\nu_3^6 dt.\label{gamma}
\end{eqnarray}
Substituting equation\,(\ref{oad}) into this equation, we find
\begin{eqnarray}
\gamma\approx 4.7\times10^{-20}\int_0^t(\nu_0+3.7n(\omega)\eta t^{3/2})^6 dt\label{gammagw}.
\end{eqnarray}
The rapid growth happens only when $\nu_0<3.7|n(\omega)|\eta t^{3/2}$, in which case
we can ignore the $\nu_0$ term in equation\,(\ref{gammagw}). Therefore,
we get
\begin{eqnarray}
\gamma=12(n\eta)^6 ({t/100\,{\rm s}})^{10}\label{gammafinal},
\end{eqnarray}
which is strongly dependent on $n$, $\eta$ and $t$, meaning the amplitude
can grow quickly with time. To reach saturation, the amplitude should grow
up to $\alpha_s/\alpha_0=10^{3}-10^8$, which means
$\gamma=\ln(\alpha_s/\alpha_0)\approx7-18$.
This implies that gravitational radiation can drive $r$-mode to
saturation in a few hundred seconds if the star accretes a significant
amount of matter. In this case, the maximum amplitude is reached when
$t_{\rm gw}+t_{\rm B,t}\approx0$, which is
$B_{\rm t,14}\approx0.24R_{11.5}^{3}M_{1.4}^2B_{15}^{-1}\nu_{3}^7$.

However, if the spin grows very slowly, the accretion torque will dominate the
amplitude growth ($|{t_{\rm gw}/ t_{\rm acc}}|>1$). Considering only the
accretion torque term in equation\,(\ref{aevo}) and using equation\,(\ref{oad}),
we approximately obtain the growth rate of the $r$-mode amplitude due to accretion,
\begin{eqnarray}
\gamma ={8.3n(\omega)t_0^{3/2}\eta[1-(t_0/t)^{2/3}]
\over \nu_c+8.3n(\omega)t_0^{3/2}\eta[1-(t_0/t)^{2/3}]},\label{aacc}
\end{eqnarray}
where $n(\omega)$ is negative and $\alpha'_0$ is the amplitude when the
accretion torque becomes dominant. In this case, the amplitude grows very
slowly with time. Then, the maximum amplitude is reached when
$t_{\rm B,t}+t_{\rm acc}\approx0$, which is $B_{\rm t,14}=-0.04
B_{15}^{-5/7}R_{11.5}^{-1/7}M_{1.4}^{13/14}\dot M_{-3}^{6/7} n(\omega)$,
where $n(\omega)<0$.

\section{Detectability of gravitational waves}

\begin{figure}
\begin{center}
\includegraphics[width=0.48\textwidth]{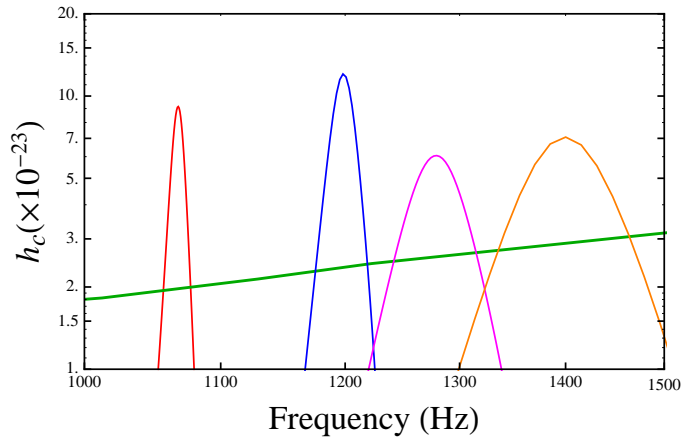}
\caption{The characteristic strain amplitude of the $r$-mode instability for
accreting magnetars locating at a distance $D=1$\,Mpc. The green curve corresponds to
ET noise $rms$ strain amplitude. The red, blue, magenta, and orange curves correspond
to the cases: $B=2\times10^{14}$\,G, $\eta=0.01$, and $t_0=3000$\,s;
$B=2\times10^{14}$\,G, $\eta=0.1$, and $t_0=300$\,s; $B=5\times10^{14}$\,G,
$\eta=1$, and $t_0=100$\,s; and $B=5\times10^{14}$\,G, $\eta=10$, and $t_0=15$\,s,
respectively.
\label{fig3}}
\end{center}
\end{figure}

When the $r$-mode is unstable in a magnetar, gravitational waves are emitted.
The $r$-mode became very interesting after gravitational waves from
the merger of double black holes were detected by aLIGO \citep{Abbott2016}.
We now test whether or not such a gravitational wave signal from a single source can
be detectable by gravitational observatories such as ET. The gravitational wave strain
amplitude for the $r$-mode, which is averaged over polarizations and orientations,
is given by \cite{Owen1998}
as\begin{eqnarray}
h=\sqrt{3\over80\pi}{\omega^2S_{22}\over D},
\end{eqnarray}
where $D$ is the distance of the source and
\begin{eqnarray}
S_{22}=\sqrt{2}{32\pi\over15}{GM\over c^5}\alpha\Omega R^3\tilde{J},
\end{eqnarray}
where $c$ is the speed of light. We mainly focus on the frequency spectrum of the
gravitational wave by using the Fourier-transform,
$\tilde{h}(f)^2=h(t)^2|{dt\over df}|$ \citep{Owen1998}, where
$f\approx 2\Omega/3\pi$ is the frequency of the gravitational wave for the $l=m=2$
current multipoles. A characteristic amplitude $h_c=h\sqrt{f^2|{dt\over df}|}$ is
usually defined and we show it in Fig.\,\ref{fig3} for different cases versus the $rms$ strain noise
$h_{ rms}=\sqrt{fS_h(f)}$  of ET, where $S_h(f)$ is
the noise strain amplitude.

The red, blue, magenta, and orange solid curves in Fig.\,\ref{fig3} are
the characteristic amplitude $h_c$ in the cases with different accretion rates
and magnetic fields, when the sources locate at a distance of $D=1$\,Mpc.
The green line is the ET noise $rms$ strain amplitude \citep{Hild2008} and the characteristic amplitudes are higher than the noise curves. Thus it is possible
to detect such sources once ET works.

\section{Discussion and Conclusion}

In this paper, we have explored the $r$-mode instability in newborn rapidly-rotating
hyper-accreting magnetars. Their spin evolution is influenced by the
accretion torque. There are two possibilities. Case I: If the neutron star is spun up significantly by accretion,
namely, the parameters must satisfy the condition of $3.7n(\omega)\eta t_0^{3/2}>794B^{2/49}_{15}
\eta^{6/49}t_0^{3/49}|n(\omega)|^{1/7}-\nu_0$, the
$r$-mode amplitude will grow quickly due to the gravitational radiation
$\ln[\alpha(t)/\alpha_0]=12(n\eta)^6 ({t/100\,{\rm s}})^{10}$. The saturation
happens when the damping due to the generation of a toroidal magnetic field
balances the gravitational radiation for
$B_{\rm t,14}\approx0.24R_{11.5}^{3}M_{1.4}^2B_{15}^{-1}\nu_{3}^7$.
The generated magnetic field reaches a value that is slightly weaker than the poloidal field.
The saturation amplitude is $\sim 0.001$.

However, if the spin-up is not significant (Case II), the accretion torque will dominate
the amplitude growth, which leads to a much slower growth (equation \ref{aacc}).
The saturation is then met when $B_{\rm t,14}=0.04
B_{15}^{-5/7}R_{11.5}^{-1/7}M_{1.4}^{13/14}\dot M_{-3}^{6/7} |n(\omega)|$.
The maximum saturation amplitude is of the order of $\sim 10^{-6} $ to $10^{-5}$. The generated
toroidal field is much weaker than that in the case of a significant spin-up.

If the magnetar is sufficiently spun-up (i.e., Case I discussed here),
the gravitational radiation from such a magnetar would be possibly detectable
with the future mission ET, if the source were located at a distance $<1$Mpc.
Core-collapse SNe and gamma-ray bursts are electromagnetic counterparts
of such gravitational wave sources. However, based on the recent estimates
\citep{Li2011,Strolger2015}, the local core-collapse supernova rate is $\sim{\rm a\, few}
\times 10^{-2}$\,Mpc$^{-3}$ per century. Thus, the detectable gravitational wave event rate with ET
is as low as roughly one per millennium.

\section*{Acknowledgements}
We thank the referee for valuable comments and constructive suggestions that have allowed us to improve
our manuscript significantly. We also thank Yi-Ming Hu, Yan Yan and Yun-Wei Yu for their
helpful discussions. This work is supported by the National Basic Research Program
(``973'' Program) of China (grant No. 2014CB845800) and the National Natural Science Foundation
of China (grant No. 11573014).

\end{document}